\documentstyle[12pt,aasms4]{article}

\slugcomment{Submitted to the {\it Astronomical Journal}}

\begin{document}

\title{Active Nuclei and Star-Forming Objects at $z>2$: 
Metallicities, Winds, and Formation Histories\altaffilmark{1}}

\author{William C. Keel\altaffilmark{2} and Wentao Wu\altaffilmark{3}}
\affil{Department of Physics and Astronomy, Box 870324, University of Alabama,
Tuscaloosa, AL 35487}

\author{Ian Waddington\altaffilmark{4}, Rogier A. Windhorst, and 
Sebastian M. Pascarelle\altaffilmark{2,5}}
\affil{Department of Physics and Astronomy, Arizona State University,
Box 1504, Tempe, AZ 85287-1504}

\altaffiltext{1}{Based on observations with the NASA/ESA 
{\it Hubble Space Telescope} obtained at the Space Telescope Science Institute,
which is operated by the Association of Universities for Research in Astronomy,
Inc., under NASA contract No. NAS5-26555.}
\altaffiltext{2}{Visiting astronomer, NASA Infrared Telescope Facility,
Mauna Kea, Hawaii}
\altaffiltext{3} {Present address: Department of Physics and Astronomy,
Louisiana State University, Baton Rouge, LA 70803}
\altaffiltext{4} {Present address: Department of Physics, University of
Bristol, H. H. Wills Physics Laboratory, Tyndall Avenue, Bristol BS8
1TL, UK}

\altaffiltext{5} {Present address: 5004 Southern Star Terrace, Columbia, MD
21044}


\begin{abstract}

We present near-infrared observations of the AGN and star-forming objects in
the field of the radio galaxy 53W002 at $z=2.39$. The star-forming objects are  
of special interest
as candidate protogalactic objects. The 1.1-2.2$\mu$m passbands sample the 
emitted optical range at this redshift, providing new diagnostics of the 
structure, metal abundance, and age of the members of this grouping originally 
selected through Lyman $\alpha$ emission. The star-forming objects are uniformly 
very blue in continuum slope, which fits with the strong Lyman $\alpha$ emission 
in indicating metal abundances that are less than half solar;  some are as 
blue as the most metal-poor local objects. They fall in a range of luminosity 
and metallicity which is not populated by local objects, indicating a shorter
star-forming history at this early epoch. The best local analogs, such as Mkn
66 and Mkn 357, either have several times lower luminosity at comparable [O/H],
or significantly higher [O/H] for comparable luminosity. 
NICMOS spectroscopy yields detections of [O III] emission
for two objects and interesting [O III] and H$\beta$ limits for the
rest, augmented by H$\alpha$ limits from IRTF imaging.
These data are satisfied by model stellar populations which have been forming
stars for the last $2-5 \times 10^6$ years before $z=2.39$. We do not see 
evidence for older
pre-existing stellar populations, either in the broadband colors or as redder
halos in which the star-forming regions are imbedded. These results suggest
that the compact star-forming objects we see at $z=2.0-2.5$ are indeed early
stages in the building of galaxies, rather than transient star-forming events
in larger pre-existing dynamical systems. The results also allow an alternative 
scheme, in which 
these are low-mass systems that are blowing winds rather than self-enriching,
in which case they should fade rapidly with cosmic epoch.
For the three prominent AGN at $z=2.39$, H$\alpha$ and [O III] emission were 
measured. Unlike the fainter star-forming objects, their line
ratios (specifically Lyman $\alpha$/H$\alpha$) show metallicities just as high
as in nearby systems. If the AGN occur in those systems which
started with the highest density and began active star formation before the
less massive surrounding objects, they will have higher metallicity (as
we see in their emitted-ultraviolet line ratios). The
``ionization cones"  seen prominently in Lyman $\alpha$ also appear in [O III]
and H$\alpha$, with a role for continuum reflection in some cases as well. 
The contrast between the AGN and fainter star-forming objects
can be broadly accomodated in a hierarchical
formation picture, though there are still important unknowns as to the
fate of the star-forming objects.

\end{abstract}


\section{INTRODUCTION}

Our view of galaxies is being revolutionized by the opening of galaxy evolution
to direct observational study. Systematic changes in the integrated
star-formation rate with redshift are well-attested, though the details of
these changes are still under debate (Madau et al. 1996,  1998; Pascarelle et
al. 1998b, Haarsma et al. 2000, Thompson et al. 2001, 
Lanzetta et al. 2002), and changes in the
galaxy content of rich clusters give the strongest evidence for evolution among
galaxies. As we move beyond this, the obvious next goal is watching galaxy 
formation, or perhaps more accurately a longer process of galaxy building,
which has repeatedly proven more elusive as we reach outward in redshift. Among
the best candidates for protogalactic objects, especially if hierarchical
CDM-like schemes apply, are the numerous faint, blue, and compact objects seen
at large redshifts. A group or cluster of these was reported at $z=2.4$ by
Pascarelle et al. (1996, P96), and similar systems prove to be common in the
Hubble Deep Fields (Mobasher et al. 1996, Marleau \& Simard 1998,  Gardner \&
Satyapal 2000).   These objects are of special interest as an early phase in
galaxy development for several reasons. Their high emitted-ultraviolet
luminosities imply correspondingly high star-formation rates, but they have no 
obvious population of counterparts in the local Universe (P96; Pascarelle et
al. 1998 = P98). Most distinguishing, they are small, with effective radii
barely resolved using HST at redshifts $z>2$, where $0\farcs 1$ translates to
$0.5-1.0$ kpc for currently popular cosmologies.  There is weak evidence that
the radial profiles are better fit by an $R^{1/4}$ bulge-like form than by
an exponential disk (P96), although composite structures are certainly
allowed. A significant fraction of these objects are closely paired, with
merging timescales (incorporating pairwise velocity differences 
200-300 km s$^{-1}$)
suggesting that they undergo rapid merging into more massive systems, or indeed
represent pieces of larger dynamical systems (Colley et al. 1997). They are
often strong Lyman $\alpha$ emitters without the additional emission lines that
are strong in AGN, supporting an interpretation as star-forming systems. These
properties  are all consistent with the idea that these objects are showing us
one of the final phases in galaxy formation, as star-forming  protogalaxies or
young galaxies grow via hierarchical merging. Their physical nature has
elicited diverse opinions (Colley et al. 1996),  hinging on whether they are
actually seen early in the process of gas-to-star  conversion, or might be
localized starbursts within substantially larger and relatively old systems,
whose surroundings are inconspicuous because of the severe
cosmological surface-brightness
dimming and the ultraviolet bias towards hot stars. Observations over a wider
wavelength range are important in resolving this issue.

The 53W002 grouping offers an especially attractive opportunity to study
several of these ``sub-galactic" objects, in the same field, as well as AGN 
of various flavors. Two of the star-forming objects in our region of interest,
objects 5 and 12, turn out to be at $z=2.05$ (Pascarelle et al. 2002), which
places their Lyman $\alpha$ so near the edge of the HST F410M passband that we
may have detected them through luck as much as statistics. We include these
object in our study since their luminosities and sizes are similar to those of
the star-forming objects at $z=2.4$. The results of the full spectroscopic
followup were not known when the NICMOS observations were specified, so we
pointed to include some objects (since found to be stars with
excesses in the F410M passband) which we would have replaced by others in
hindsight.

We present here a near-infrared study of the region surrounding 53W002, using
HST NICMOS and IRTF data. This wavelength regime shows us the emitted optical
range at this redshift, so we expect to see stars cooler than the B-type stars 
that dominate the emitted ultraviolet range originally used to find these 
objects. The full spectral
shape from 1100--6600 \AA\   in the emitted frame gives a much more sensitive
way to see how old the stellar populations are and provide evidence of internal
extinction, while spectroscopy in the near-IR can also indicate the gas
abundances from [O III] and Balmer emission. The Lyman $\alpha$/H$\alpha$ ratio
can tell us, albeit in a complicated way, about the abundances and geometry in
the gas and dust. For the three luminous AGN in this sample, we can compare any
host galaxy properties we can measure, and examine their extended ionized gas
to see how well an ionization-cone model (as suggested by the Lyman $\alpha$ 
structures) explains them. We analyze HST NICMOS
images and grism spectroscopy, and IRTF imaging in the $K$ band and redshifted
H$\alpha$, to refine our understanding of how old and how large these putative
young galaxies really are.

\section{OBSERVATIONS}

\subsection{NICMOS Imaging and Spectroscopy}

Two overlapping fields in the 53W002 region were observed with NICMOS, using
the widest-field NIC3 camera during the first NICMOS 
refocus campaign in January 1998.
Direct images of 2439 seconds' duration  were obtained in the F110W and F160W
filters (corresponding roughly to emitted central wavelengths 3200 and 4700
\AA\  at $z=2.4$), along with a similar exposure in F175W (emitted wavelength
5100 \AA\ ) to support the grism spectroscopy, although the thermal background
in this passband is substantially higher than in F160W, and its S/N ratio 
correspondingly worse. A single image set in F240M was also obtained, but the
actual thermal performance of the telescope meant that only the few brightest
objects in the field were detected.

We also obtained slitless spectroscopy with the NICMOS G141M grism for these
two fields, with a total exposure time of 8054 seconds for each field. The
dispersion was fortuitously oriented nearly north-south, perpendicular to the
extended emission-line structure of 53W002, giving us a good spatial trace of
[O III] $\lambda 5007$ in its ionization cones.

To complement these data for 53W002 itself, we obtained a single-orbit
exposure at higher resolution using the NIC2 camera and F160W filter. The radio
core-jet or small double structure of 53W002 makes it the best candidate for a
dynamically relaxed, perhaps young elliptical galaxy in this grouping
(Windhorst et al. 1998), so we
use these data to look for any redder, extended de Vaucouleurs profile, as
well as the role of line emission as evaluated with the help of the grism
spectrum. At $z=2.39$, [O III] $\lambda 5007$ is observed at 1.697$\mu$m, so the
two [O III] lines are strong contributors in the F160W passband. 

The two NICMOS fields are outlined in Fig. 1 on a WFPC2 $I$-band (F814W)
image showing the
spectroscopically confirmed high-redshift objects, augmented  with a
ground-based $V$ image (Keel et al. 1999) to fill in regions outside the WFPC2
footprint. The NICMOS fields include all three of the AGN found in P96 as well
as 6 of the fainter Lyman $\alpha$ emitters (not all of which were detected in
the NICMOS images). Objects 10 and 13 of P96 have since proven to be stars
(Pascarelle et al. 2002), which we suspected originally given their red colors, 
so they are  not counted henceforth.
For reference, Table 1 lists the targeted objects and their
spectroscopic redshifts, including Lyman $\alpha$ results from Keck
multislit observations presented by Pascarelle et al. (2002). The
redshift of object 1 is ambiguous, with their paper listing $z=3.05$ as
a possible redshift based on a marginal emission feature at 4955 \AA\ .
We also consider its properties if it in fact lies at $z=2.4$,
in which case its actual line emission would be weaker and unseen
in their spectrum.

We processed the NICMOS images to remove the variable ``pedestal"
effect, which is a change in the bias level between the quadrants of the detector. Our
fields are sparse enough that blank sky could be used to assess the offsets
quite accurately, which was done by simultaneously minimizing flat-field
residuals in blank-sky regions and requiring continuity of the mean sky level
across the four quadrant seams. Standard STScI/NICMOS library flat fields were
used.  Each final image is the clipped mean of a $3 \times 3$ dither pattern 
with
spacing $1 \farcs 67$ (8.3 pixels), done in three successive strips of three exposures
each, to allow an exact match between coordinate systems in direct and grism
images while retaining fine control over execution time. Pipeline removal of
cosmic-ray events from the MULTIACCUM readout sequences was
very effective, so that no further cosmic-ray  treatment was required. 

Close inspection shows ghost images 8 pixels along the detector $+x$ axis from
the three bright stars in the NIC3 fields. These are apparently image
persistence artifacts, as propagated through mosaicking with the 8-pixel
dither step between successive observation sequences. None falls  in a
particularly damaging location for our purposes. In the grism images, there are
several ``zero-order" images of bright objects in and beyond the southern part
of each field observed, which could potentially be confused with genuine 
(high-redshift) emission-line objects.

The dither sequence, on a set of (integer+1/3)-pixel centers, allowed drizzle
reconstruction to improve sampling. We produced a set of F110W and F160W
drizzled images on a $0 \farcs 10$ grid, and show the F160W images in Fig. 2. Photometry
was carried out both using SExtractor to estimate total fluxes, and by
explicit aperture photometry with curve-of-growth  corrections based on the
PSF. Comparison shows that these almost always agree at better than the 15\%
level, with a few larger discrepancies traced to resolved structures such as the
extended emission-line regions around the AGN (which are prominent in F160W due
to [O III] emission appearing at 1.70$\mu$m). This dispersion due to image
structure is larger than the purely statistical error in the photometry, so we
take the actual photometric error as 15\% except for the two objects fainter
than AB magnitude 24.7 in F110W, in which the internal statistical error
dominates (giving errors of 0.2--0.3  magnitude). The grism data were
processed using the NICMOSLOOK routine, to produce a synthetic flat field at
each spectral pixel, based on the wavelength of the dispersed light from the
associated object as identified from the direct image, and subsequent
extraction of the  spectra with windows optimized for the signal-to-noise ratio.
A synthetic flat field was generated for each extracted spectrum,
interpolating between narrow-band flats to derive the sensitivity appropriate
to the wavelength corresponding to each pixel for the object's position. The
filters used to generate these flats are concentrated near 1.1, 1.6, and
1.9$\mu$m, and the filter curves suggest that linear interpolation between flats
may generate fractional errors of 10\% in the flat-field structure at the
least-constrained wavelengths near 1.3$\mu$m. However, the 
wavelength region of redshifted
H$\beta$ and  [O III] $\lambda 5007$ (from 1.65-1.70$\mu$m), is well constrained
by this process, so that flat-field errors in this spectral range are minimal. 
For
emission-line objects too faint to have their spectra automatically extracted,
we sought emission lines by examining the grism images at the $x,y$ locations
given by the object's coordinates in direct images and the predicted wavelength
of [O III] $\lambda 5007$, based on the dispersion relation for the
G141 disperser from the NICMOS Data Handbook (v3.0). The tilt angle between the
dispersion axis and detector rows can change with time, and was derived from 
the data themselves using spectra of bright stars. In two cases we have
formal detections of [O III] line emission from these faint Lyman $\alpha$ 
emitters, that are significant above the 2$\sigma$ level.

\subsection{Ground-based 2$\mu$m imaging}

We used the 3-m NASA Infrared Telescope Facility in July 1996 and April 1997
to observe the 53W002 region in $K$ and in redshifted H$\alpha$, with the 
$256 \times 256$-pixel NSFCAM array configured for $0 \farcs 30$ pixels. The H$\alpha$
data used a circular variable filter (CVF) tuned to 2.22$\mu$m. We measured the
filter bandwidth to be FWHM=0.040$\mu$m  ($\Delta z = 0.05$ for redshifted
H$\alpha$) from a spectral scan of  NGC 7027 in Br$\gamma$ and the adjacent
H$_2$ S(1) line (at 2.166 and 2.247 $\mu$m, respectively).  Individual $K$
exposures were 10 seconds, coadded in groups of six before saving to disk, 
while
exposures through the CVF were 60 seconds long. From two observing runs, we
obtained total integrations of 3.9 and 10.9 hours in the $K$ filter and CVF,
respectively.
The effective integration time per pixel varies substantially across the field,
due to both short- and long-timescale dithering. Flux calibration used UKIRT
faint standard stars (Hawarden et al. 2001).

To account for changes in the near-IR atmospheric emission spectrum on
timescales of hours, we constructed sky-frames from sliding clipped means of
data subsets. The seeing varied widely during the single night of 1996
observations, so we rejected images with poor image quality as assessed from
bright field stars. The 1997 data suffered from a slowly time-variable PSF due
to changes in the mirror figure with hour angle, so they were stacked into
approximately one-hour blocks coinciding with changes in filter or large-scale
dithering. The region of 53W002 itself was covered mostly in the 1996 data,
while the area including the Ly$\alpha$ emitters to its west was included in
both data sets, so we analyze these objects from the overall sum. For pure
continuum objects, the count-rate ratio between narrow-band and $K$-band images 
is
consistent within 2\% over the range of colors encountered in this field, so we
can use this ratio as a straightforward indicator of line emission within the 
CVF passband.
The three AGN (53W002 itself and objects 18 and 19 of P96) are
strong H$\alpha$+[N II] sources, with spatially
integrated line fluxes from the NICMOS and IRTF observations
as listed in Table 2. The table includes equivalent widths in \AA\  in the 
emitted frame; the
nearby narrow-line radio galaxy Cygnus A is included for comparison, from data 
in Owen, O'Dea, \& Keel 1990 and Osterbrock \& Miller 1975. The H$\alpha$ 
emission from 53W002 is spatially resolved, elongated E-W
(Fig. 3), roughly matching the structure seen for Lyman $\alpha$
in WFPC2 imaging and for [O III] in the NICMOS 
spectrum.

We detected none of the $z=2.4$ star-forming objects in H$\alpha$+[N II] line 
emission, giving lower limits to the
Ly$\alpha$/H$\alpha$ ratio. The detection thresholds vary somewhat with
position, due to different effective integration times after mosaicking the
various sets of pointings. Typical limits on H$\alpha$+[N II] flux are around
$1.1 \times 10^{-16}$ erg cm$^{-2}$ s$^{-1}$.

\section{SPECTRAL ENERGY DISTRIBUTIONS}

We combine the integrated fluxes from the NICMOS data with new
measurements from the earlier WFPC2
observations (P96, P98) to define the broadband spectral shape of the
star-forming objects over the 1100-6500 \AA\  emitted range. The WFPC2
fluxes have been improved through ``on-the-fly" reprocessing with improved
flat fields and bad-pixel maps, as compared with the earlier analysis.
The relevant
fluxes are given in Table 3, listed as the equivalent flux density at
band center for a flat spectrum. For brevity we denote measures through WFPC2
filters by their nearest standard photometric equivalents: $B$ (F450W), $V$
(F606W), $I$ (F814W).


As shown in Fig. 4, the objects are quite blue across this entire spectral
range, immediately suggesting that the stellar populations are both young and
metal-poor (since such a steep slope leaves little scope for dust reddening).
The three spectroscopically identified AGN distinguish themselves from the
putative star-forming objects not only in flux but in having substantially
flatter spectra in $F_\lambda$. To investigate the age and allowed metallicities
of the star-forming objects, we follow the precepts of Calzetti et al. (1994)
in using the UV slope parameter $\beta$ (where $F_\lambda \propto 
\lambda^{\beta}$ as fitted across the 1200-2600 \AA\  range). This was 
derived by least-squares fitting in the log
domain (Table 4). Many of the derived errors cluster closely, because much 
of the 
overall photometric error comes from either possible scale errors or
aperture effects, and is thus not a very strong function of flux.

Calzetti et al. find that the intrinsic spectral slope of model populations in
the extreme cases of an instantaneous burst (less than  $2 \times 10^7$ years
old, to keep substantial ionizing flux), and continuous star formation at a
constant rate, differ by no more than 13\% in slope about the $\beta = -2$
reference point over the 1200-2600 \AA\  span of good IUE spectra, so that 
the observed slope of a
star-forming system can be used to estimate the effective reddening. Using
large-aperture integrated spectra, so that we deal with scattering 
and mixing  of multiple
star-forming regions in a way appropriate for comparison with the integrated
properties we measure for the high-redshift objects, they show that nearby
systems follow a correlation between slope of the UV spectral energy
distribution (SED) and emission-line abundances, for a plausible
reddening-metallicity correlation. Our slope measurements, being based on a few
broad-band data points, are necessarily less sophisticated than the targeted
spectral windows used by Calzetti et al.
As noted by Bell et al. (2002), the difference in $\beta$ as derived
from fitting in the purest continuum windows and from images in these
broad passbands gives a typical error $\pm 0.2$, comparable to our
purely observational statistical errors. We derive values
$\beta = -1.1$ to -2.3 from fits across the entire observed
range from 0.45-1.6 $\mu$m for the star-forming objects in the 53W002 grouping 
(Table 4). Slopes over subsets of this spectral range are consistent, but
give larger internal errors (and scatter), ranging from -1.4 to -2.4 between
0.55 and 1.1 $\mu$m. Only a few nearby star-forming objects have continua this
blue; Calzetti et al. (1994)  list only NGC 1705, UGC 9560, and Tol 1924-416
having $\beta < -1.8$, while Meurer et al. (1999) add seven more (NGC 4861 and
Mkn 153 at $\beta < -2.0$ plus Mkn 66, NGC 3991, NGC 3738, UGCA410 and Mkn 357
with $-1.8 > \beta > -2.0$). All these are classified as blue compact galaxies
rather than within the ordinary Hubble types. THree of the six 
objects in the 53W002 field fall in this regime.

Although the correlation between spectroscopically derived [O/H] and $\beta$
has substantial scatter ($\pm 0.2$ in log (O/H) at 12+log (O/H)=8.3, with the
scatter greatest at low metallicity), the only nearby systems
as blue as objects 1, 12, and 60 in this sample 
have 12+log (O/H) $<$ 8.5, which implies less than half solar (12+log
(O/H)=8.82).  The extreme low-abundance dwarf I Zw 18 falls in this range, with
$\beta = -2.43$ from  Meurer et al. (1999) and 12+ log (O/H)=7.18 from Izotov
\& Thuan (1999). The color-abundance relation clearly flattens at such low
values, but even so, the metal abundances implied by the blue colors
are low enough to suggest that we are seeing genuinely lower abundances
at $z=2.4$ than are found in the most similar systems today. This is
illustrated in Fig. 5 by both the observed and dereddened UV luminosities
of the 53W002 star-forming objects, most of which fall
outside the range of luminosity and $\beta$ populated by local objects.

Low metallicities for these systems
are especially interesting in view of the high UV
luminosities and derived SFR. In a simple
picture of starbursts in an interstellar medium, the metallicity of the gas may
reflect a combination of pre-burst conditions and metals expelled from the
first supernovae of the starburst population itself. One might then expect a
given starburst to have greater metallicity in the present Universe, after many
Gyr of previous star formation has enriched the ISM, compared to the early 
Universe,
in which the burst we observe
may mark the first significant episode of star formation.
Enrichment to about the SMC level should be quite rapid, on the
order of $10^7$ years, unless the enriched material is quickly
mixed with large amounts of pristine gas (Izotov \& Thuan 1999, Kobulnicky 
1997). 
The general luminosity-metallicity relation can be examined for differences in
locus between local and high-redshift star-forming objects. To do this, we 
compare metallicities with
UV luminosities at emitted wavelengths 1100--1300 \AA\ , for which IUE and HST
data are available on nearby objects (from Calzetti et al. 1994 and Meurer et
al. 1999, augmented by [O/H] measures for I Zw 18 and NGC 4861 from Thuan \&
Izotov 1999). We consider sets of cosmological parameters with  $\rm H_0 =
60-80$ km s$^{-1}$ Mpc$^{-1}$, and both $\Omega_M=1$ and $\Omega_M=0.3,
\Omega_\Lambda=0.7$. These combinations, at $z=2.4$, give ranges of linear scale from
4.9--9.5 kpc arcsec$^{-1}$, and luminosity distance $D_L=11.6-22.6$ Gpc, at 
lookback times
6.8-12.5 Gyr and cosmic ages 1.2--3.1 Gyr. Thus for these high-redshift objects 
the luminosities we derive will vary by a factor 4 depending just on
the adopted cosmology; in Table 5 we quote the geometric mean of these 
extremes. Smaller
$\rm H _0$ and larger $\Lambda$ increase the derived luminosity. The
star-forming  objects in the NICMOS fields all fall within the range $1.5-9
\times 10^{40}$ erg s$^{-1}$ \AA$^{-1}$ in this (emitted) passband, quite
luminous indeed. The only object well within this luminosity range in the IUE sample
used to derive the local relationships is Mkn 357 ($z=0.05$, $L_{1200}=5.7
\times 10^{40}$). A metallicity-luminosity relation appears in the data for
12 + log (O/H)$<8.6$, above which extinction corrections become large enough that the
star-forming regions are dimmer when using uncorrected data. The important point
is that the objects at $z>2$ occupy a region in this plane which contains no
nearby examples (Fig. 5).

The most luminous nearby object with 12 + log (O/H)$<8.5$ is Mkn 66, at 8.39 
and L(UV)=$4\times 10^{39}$ erg s$^{-1}$ \AA$^{-1}$, which is 4-20 times less
luminous than the high-redshift objects  (so there is no overlap for any of the
cosmologies we consider). The gap grows for lower [O/H] values. Correction for
internal absorption, which we do following Meurer et al. (1999) in taking
$dA_{1600}$/d $\beta$ = 1.99, doesn't change this situation, though it does
increase some luminosities by an order of magnitude (Fig. 5).

%
%
%
%

These stellar populations cannot be extremely young bursts (a few $10^6$ years),
or the optical line equivalent widths would be very high against the combined
nebular and stellar continuum (as in Mkn 490, observed by De Robertis \& 
Osterbrock 1986); the equivalent width of H$\beta$ saturates against 
the recombination continuum at 1350 \AA. The continuity between WFPC2 and
NICMOS fluxes, and the comparably good fit of the power law across
the whole observed range, shows that the emerging light is dominated
by young populations. We see no clear evidence of any upturn at longer
wavelengths which would come from a significant redder and older
population (section 6). Thus the populations do not furnish us the leverage 
needed for a detailed fit to models of various ages or mixed populations
involving multiple star-forming  episodes.

\section{EMISSION LINES IN THE STAR-FORMING OBJECTS}

Examination of individual pixels around the predicted location of the
[O III] lines shows detections of $\lambda 5007$  (2-4 $\sigma$ at 
independently specified  wavelengths) in two of the star-forming objects 
which are brightest at 1.6$\mu$m, object 11 at $z=2.451$ and object 12 at 
$z=2.05$. This procedure gave higher signal-to-noise ratio than the
standard NICMOSLOOK spectral extraction for detecting faint and narrow 
emission features, whose results showed no significant line emission in either
H$\beta$ or [O III].
Crosscuts through the grism images, showing emission at the expected wavelengths 
of [O III], are
shown in Fig. 6. Both these objects show double morphologies in WFPC2 F450W
images, which might either represent merging or multiple star-forming regions
in a single galactic potential. In combination with the Ly$\alpha$ fluxes and
SED slopes, even our upper limits for lines redshifted into the near-IR set
interesting constraints on the effective ages and metallicities of these
objects. The measurements and upper limits are given in Table 6. For comparison,
upper limits to H$\alpha$ emission for these objects (Section 2.2) are
typically $7 \times 10^{-17}$ erg cm$^{-2}$ s$^{-1}$. This already indicates a
minimal role for dust, since the Lyman $\alpha$/H$\alpha$ ratio is required to
be larger than encountered in nearby star-forming systems (e.g.
Giavalisco, Koratkar, \& Calzetti 1996); lower limits
for the ratio range from 0.3--3 for various objects. The H$\alpha$ limits imply
upper limits on H$\beta$ slightly more stringent than do the NICMOS spectra
themselves, typically $2 \times 10^{-17}$ erg cm$^{-2}$ s$^{-1}$, for those 
objects
at $z=2.39 \pm 0.013$ (including, as it happens, neither of the objects with
possible  [O III] detections, since their redshifts fall outside the range for
{\it bona fide} group members for which the narrow CVF
H$\alpha$ filter was tuned).

The continuum slopes make it clear that these objects must be in the
low-metallicity regime, so they are likely to fall on the lower branch of the
relation between [O III]/H$\beta$ and [O/H], often used to estimate
metallicities for faint objects by the ``bright-line" method
(introduced by Pagel et al. 1979). The line limits (and possible 
[O III] detections) can put combined constraints on the duration of star formation
and metallicity. Significant star formation is still in progress as we see these
objects, since there is enough ionizing radiation to power the Lyman $\alpha$
emission, although radiative-transfer effects mean that we cannot use
its intensity by itself as a reliable estimator of the rate of star formation. 
The equivalent width of H$\beta$ was considered as an age indicator
by Copetti et al. (1986). We use the more recent models presented by Stasinska
and Leitherer (1996). Effective burst ages in the range $(3-8) \times 10^6$ years
satisfy the equivalent-width limits for the emitted-optical lines while still
having strong Lyman $\alpha$. The H$\beta$ limits rule out very young,
high-ionization starbursts; there must be enough cooler stars to contribute 
diluting light to the continuum around H$\beta$. Photospheric absorption
in the stellar populations will be small compared to these limits;
the largest equivalent widths seen in main-sequence stars are 13.5 \AA\ ,
from the data presented by Jacoby, Hunter, \& Christian (1984), while
the largest values for model populations with a Salpeter-like initial-mass
function are only 4 \AA\ .

Stasinska and Leitherer note
that there is a small spread in model parameters needed to reproduce the
observed properties of low-metallicity starbursts, so that the [O III]
equivalent width is also useful as an age indicator. Using their models for
metallicity 0.25 and 0.1 solar, the H$\beta$ limits imply equivalent burst ages
greater than $2.5 \times 10^6$ years  for the high limits, and $4  \times 10^6$
years for the lowest limit (object 5). For the two nominal detections of
[O III], its equivalent width gives broadly concordant ages, $(2-5) \times 10^6$
years for the relevant metallicity range.   If these detections
are valid, they put lower limits on the gas metallicity, since the implied 
[O III]/H$\beta > 3$ for these two objects. However, again following the models
published by Stasinska  \& Leitherer (1996), these limits only imply that 
the abundances are less than solar, since the predicted line ratios (in the age
range matching the equivalent widths) are high for all the abundances they
considered. This is much less stringent than limits from the continuum slope,
though refinement from ground-based data is now possible using IR spectrometers
on 8-10m telescopes. We note that the two objects with [O III] detections
are the reddest and presumably most metal-rich of the NICMOS sample,
which fits with the picture that these are all on the low-metallicity
branch of the relation between O/H and [O III]/H$\beta$, so that
lower O/H in fact weakens [O III] emission.

There is a dual paradox in the combination of high rates of star formation
needed for the observed ultraviolet luminosity, low metallicity, and
number of objects seen in the small volume of the 53W002 grouping.
If these are continually forming stars over a long time, we should see
low abundances in a closed-box system for only a very short time,
and might expect to see a much larger number of older systems with
redder colors and higher metallicities, which the source counts in this
field do not allow. If indeed they have all begun star formation only
$\approx 10^7$ years before our view, it seems unusual good fortune
that they have done so nearly simultaneously so that we see so many
at once. We might thus consider pictures in which such systems
might be observable with these properties for longer times, and ask what
kinds of present-day galaxies they might become.

Episodic star formation makes some of these issues more tractable,
though still suggesting that there are more similar objects in a dim
state by a factor related to the duty cycle of active star formation.
Such episodes might be triggered by the equivalent of mergers (though
now for systems which might or might not already be considered
distinct galaxies, and might not yet have undergone important
levels of star formation). This notion has some support from the
fact that half of star-forming objects in our sample have double
structure on 2-5 kpc scales, and the well-documented relation between
tidal interactions and star formation in the present Universe.
This mechanism lengthens the timespan over which the objects are
(sometimes) bright, but does not address the low-metallicity issue.

There is substantial evidence that global winds (superwinds) were important
during the early phases of galaxy evolution. Powerful winds are
suggested for high-redshift galaxies by strong and slightly
redshifted Lyman $\alpha$ emission (e.g. Pettini et al. 2001),
by the amount of early enrichment needed to account for abundances
in the intracluster and intergalactic media (Tripp, Savage, \& Jenkins 2000),
and by a connection between strong Lyman $\alpha$ and global winds for
starburst galaxies (Keel 2002). At low redshifts, the fossil record of
dwarf galaxies and ongoing winds in starbursting dwarfs suggest that
many low-mass galaxies have been swept clean by winds rather than
being able to self-enrich with processed material for long times
(Marlowe et al. 1995).

Winds from the star-forming objects we see at $z > 2$ would help explain
both the strong Lyman $\alpha$ emission and low abundances, since processed
gas would be removed as fast as it is produced. If this only happens
once, these objects are not the progenitors of any obvious present-day
galaxy population, but multiple cycles of gas inflow and outflow
could yield an interesting stellar mass and slow buildup of metals.
Such cycles could ameliorate the fading problem, giving a final
stellar mass that corresponds at least to current dwarf systems.
Our direct limits on the star-forming histories of these objects
are not very strong (section 6), although we can say that they
allow a mass in stars (for a Salpeter-like initial mass function) which
is an order of magnitude greater than the mass of stars with age
less than $10^7$ years, so that in principle many similar episodes
could have occurred.

\section{STRUCTURE AROUND AGN IN THE EMISSION LINES AND CONTINUUM}

Observations in the emitted UV have shown extended structure around all three
bright AGN in this field, as seen in Lyman $\alpha$ (Windhorst et al.  1998,
Keel et al. 1999), an interestingly high fraction even if one based on  small
numbers. The brightest parts of the extended emission are elongated, and
in two cases roughly triangular with the nucleus at the apex. In line
with observations of many other AGN, such structures may
fairly be interpreted as ionization or scattering cones, an idea
we explore using these data. Further information on these
features might distinguish the roles of scattering and {\it in
situ} line emission, and to tell whether this material is part of outflows
driven by the AGN or might represent ambient (enriched) material. The near-IR
data show these structures as traced by H$\alpha$ and  [O III] $\lambda$
$\lambda$ 4959, 5007, adding not only new ions but new combinations of spatial
and spectral resolution. The line intensities and equivalent widths are given
in Table 2, along with Lyman $\alpha$ fluxes from Pascarelle et al. (2002)
for comparison. The NICMOS spectra are shown in Fig. 7. At this redshift, 
[O II] $\lambda 3727$ falls in a region of poor throughput, so it was not 
clearly detected in any of these AGN.

The Lyman $\alpha$/H$\alpha$ ratios for the three AGN range from 0.2 (53W002)
to 2.0 (object 19). Values near 2 are usual for nearby narrow-line radio
galaxies over a wide range of ionization (Keel \& Windhorst 1991). H$\beta$ is
not clearly detected in any of these objects. It would appear 6 pixels blueward
of the $\lambda 5007$ peak, but the two objects with strongest [O III] show
spatially extended emission such that the monochromatic [O III] emission
overlaps the location of H$\beta$. A multicomponent fit to the profile in
53W002 gives a formal value H$\beta$/[O III]=0.06, best taken as an upper
limit. Ratios this large are common in narrow-line AGN, such as Cyg A
(Costero \& Osterbrock 1977).

We note that the emitted-ultraviolet spectra (Pascarelle et al. 2002)
of these three AGN,
like essentially all others where the spectrum is not compromised
by broad absorption features, show solar or supersolar metallicities
for the broad-line regions.
Hamann et al. (2002) find that the nitrogen abundance is most
robustly measured from strong UV lines, especially $\lambda 1240$,
and that formation of massive stars must have been active for $\approx 10^8$ 
years to yield such enrichment. This is in stark contrast to the
low abundances derived from colors and line-ratio limits in the
star-forming narrow Lyman $\alpha$ sources. 

The colors and emission-line structures of the extended emission around
the AGN have suggested a mix of mechanisms for their radiation.
The strong and very extended Lyman $\alpha$ cloud around object
18 (Pascarelle et al. 1996, Keel et al. 1999) requires {\it in situ}
photoionization, probably by an anisotropic continuum source. However,
the NICMOS spectrum shows that the continuum at 1.5-1.6$\mu$m
is spatially
resolved, spanning a $1 \farcs 4$ range, suggesting that scattering is also
important (whether or not we have a direct view of the nucleus). 
The nature of the putative scattering medium is important for understanding 
the history of star formation in the AGN hosts. If dust is responsible,
we have very different conditions in the AGN host galaxies, which are
metal-rich enough to be producing substantial dust for transport over 
tens of kpc, while their non-AGN neighbors are so metal-poor that
we see only mild evidence of internal extinction in the ultraviolet. This 
correlation between derived metallicity and presence of an AGN suggests
that the low-luminosity star-forming objects have not yet hosted a powerful AGN. 
The extent of transport of the dust and enriched gas might be important
in some cases if it can interact with the nearby star-forming systems.

\subsection{Reflection nebulosity in Object 18}

Object 18 has complex extended structure in Lyman $\alpha$ emission
(Pascarelle et al. 1996, Keel et al. 1999). The roles of ionization
and reflection are not well separated by earlier imaging data, and the grism
spectrum provides new evidence that reflection (presumably from
dust many kiloparsecs from the nucleus) plays a significant role.
A crosscut through the grism data perpendicular to the dispersion
(which is almost north-south) shows that its image is resolved and
distinctly double. Such resolution might be due to a genuinely
complex starlight distribution or to scattering of radiation from the
AGN. A contribution from scattering would be especially interesting, 
implying that grains are abundant
and widely distributed (perhaps by global AGN-driven winds). All three
AGN in the field we observed show extended emission-line clouds at
lower levels, and it is at least suggestive that these are objects
with higher abundances as judged from the AGN emission lines. 

A combination of scattered and {\it in situ} line emission
fits with the Keck optical spectroscopy from Pascarelle et al. (2002).
A slit through the extended emission about 1" NE of the core of object 18
shows strong line emission, including the C IV $\lambda 1549$ feature
(a blended doublet) which is
characteristic of the dense environment in active nuclei, seen in the
cloud at a level C IV / Ly $\alpha$=0.17. The ratio is somewhat
larger in the core, at 0.30, but it is not obvious that we can derive
relative contributions of scattered AGN light and {\it in situ}
photoionization from these two ratios, since we do not know  whether
the core spectrum gives a direct view of the broad-line region.

The grism data furnish what amounts to a monochromatic image in [O III]
once the smooth continuum has been subtracted (shown in Fig. 8). Two prominent
emission lumps appear, roughly coincident with the AGN core and the
Lyman $\alpha$ cloud $0 \farcs 6$ to its northeast as seen in the WFPC2 data.
The core:cloud flux ratio is 0.6 in [O III], substantially higher
than in the continum. The mix of scattered to local radiation must
be different on the two lines of sight; lacking confirmation of a
direct view of the core, these values cannot yet specify this mix.

\subsection{Extended line emission and the host galaxy of 53W002}

Given the brightness of possibly scattered light in object 18, and since
object 19 is nearly a point source, 53W002 itself is the only
one of the AGN whose host galaxy we can examine in detail. Even 
in this case, the imaging analysis has not been altogether straightforward
in the presence of extended emission-line structure along the
radio-source axis, which might be
accompanied by continuum reflection as well (Windhorst et al. 1998).
Considerable new information is added by the {\it Subaru} spectrum
of 53W002 across the JHK bands presented by Motohara et al. (2001).
They detect a substantial Balmer jump in the starlight, rather than
the 4000-\AA\  break which would be strong in populations much more
than about $10^9$ years old. This spectroscopic analysis avoids
some of the uncertainties in using the emitted-UV colors of the galaxy after
removing various estimates of the contribution from the AGN and
associated extended structure (as was done in W91, W94, and W98).
In all cases the age derived from the SED is less than $5 \times 10^8$ years.

The NIC2 F160W image and slitless spectrum can be combined to improve
our separation of line emission and continuum structure in the host
of 53W002. At the low dispersion of the G140L grism, the two
[O III] lines are blended by the resolved emission-line structure, so that
subtracting an interpolated continuum gives a good approximation to
a monochromatic [O III] image (Fig. 8). The [O III] emission comes largely from
two regions $1 \farcs 0$ apart situated symmetrically about the nucleus along the
same axis as the Lyman $\alpha$ extent, consistent with following
the radio-source axis. Line emission from the AGN core itself is
rather weak, about 15\% of the total flux, and the two emission regions are
unequal with almost 2/3 of the total flux coming from the western one.

We can then rotate this [O III] image,
scale properly in both pixel size and intensity, and subtract it
from the NIC2 broadband image to yield an improved image of the
galaxy's continuum structure alone (including the nucleus and any scattered
AGN light). This passband, centered near 4800 \AA\  in the emitted
frame, should be a better tracer of its overall stellar structure than the
emitted-UV images from WFPC2. The result shows that much of the
asymmetric structure in F160W, which roughly coincides with the
[O III] emission, is not line contamination, but represents continuum light.
The line emission accounts for about 16\% of the total radiation
in this passband, and its subtraction (based on the lower-resolution
grism data) does not change the overall morphology in this respect.
This reinforces the conclusions from Windhorst et al. (1998), that
there is a substantial contribution by reflection (or local, triggered
star formation) along the radio axis.

\section{OLDER HALOS AROUND STAR-FORMING SYSTEMS?}

The difficulty of detecting typical galaxies at $z>2$, and difficulties in
matching the correlation statistics of galaxies in deep fields
to dynamical-friction behavior, have led to suggestions that 
many of the faint objects seen in deep fields do not represent
distinct galaxies, but star-forming regions imbedded within larger
dynamical entitities (Colley et al. 1996, 1997). This
issue is important not only for interpreting the number counts,
but for our whole picture of galaxy building. We can test for the
role of surface-brightness and ultraviolet favoritism in our
sample by seeking any redder halos around the star-forming objects,
and using color information to tell whether the apparent double
objects might represent associations of a galactic center
(presumably with a longer history of star formation) and an
offset region of active star formation but less stellar mass.
Redder halos are common among the rapidly star-forming systems
such as blue compact galaxies (Cair{\' o}s et al. 2001),
in accord with studies of the stellar populations in some
local star-forming galaxies (Loose \& Thuan 1986,
Schulte-Ladbeck et al. 1999a,b and references therein).

To make color-index images of the Lyman $\alpha$ emitters,
we convolved the WFPC2 $B$ images to match the resolution of the NICMOS
data, at both F110W for better resolution and F160W to sample
longward of the 4000-\AA\  break.
Due to the boxy effective PSF resulting from drizzling the
NICMOS images, and the
different directions of diffraction structures in the 
two data sets, a differential point-spread function was
constructed through deconvolution of a star seen in the NICMOS data
with the same star from the WFPC2 data, after resampling them to the
same pixel scale. The differential PSF had significant
negative sidelobes, because of the diffraction spikes, and was slightly 
tweaked by a further Gaussian
blur of 0.5-pixel ($0\farcs 05$) FWHM to give neutral color gradients in
the outer edges of images of bright unresolved objects.  

We find no evidence of redder halos in the NICMOS data; there is a slight
trend in the opposite sense, to have bluer halos, although the
PSF matching is sensitive enough to numerical details that
we can't claim that this tendency is real. Specific flux
limits depend on how extensive the underlying redder population is
taken to be. Limits to the overall amount of light from a redder
population can come from the photometry. Since the SEDs are well
described by the power-law forms which are empirically good
fits for local star-forming systems over a wide range of metallicity,
an upper limit to the amount of light from a redder population can come
from considering the mean residuals at the longest wavelengths
with respect to a power law fit at shorter wavelengths. We
consider excesses measured at 1.6$\mu$m above a power-law $\beta$ fit to
data from 0.45-1.1$\mu$m, which will be especially sensitive to
populations old enough for either a strong Balmer or 4000-\AA\  
break. Of six objects, there are equal numbers with positive and
negative residuals at 1.6$\mu$m with respect to the power law
fitted to shorter wavelengths. The mean value is -0.01 dex with
a standard deviation of 0.20 dex. Using the standard deviation of
the mean, no more than 0.08 dex (or 20\%) of the flux
emitted at 4800 \AA\  comes from an older population.
Any old population much brighter than this would have to be so extended,
and hence of such low surface brightness, that it is not reflected in
apertures more than 10 times the effective radius of the UV component.

The properties of the three Lyman $\alpha$ emitters which show double
morphologies in the WFPC2 images are also relevant to the question of
whether we are really seeing the whole systems here, or only star-forming
pieces. Given the color and surface-brightness selection biases inherent
in optical detection at large redshifts, these could equally well be
genuine, dynamically multiple systems, or represent a nucleus and
offset star-forming region. The latter case would generally give
different colors for the two components, especially when observed
at redder wavelengths where any pre-existing population in the nucleus
would be more prominent. The three such putative paired objects in the
NICMOS fields are too close together to be clearly resolved in the
NIC3 data (Fig. 9), but we can check for significant color differences between
components by measuring shifts in peaks or centroids between the WFPC2 and 
NICMOS
data when measured at a common resolution. Any such shifts are below
our error levels, which are about 0.2 drizzled pixel or $0\farcs 02$
as measured from bright stellar images. This implies that the flux ratios 
between components
in no case change by more than about 50\% from $B$ to 1.6$\mu$m.
Specifically, the northern component of object 12 is about 0.3 magnitude
brighter at 1.6$\mu$m than at $B$ relative to its close companion 
$0 \farcs 27$
away. Object 11 has components $0 \farcs 4$ apart, just resolved in the
NICMOS drizzled images. In this case as well, the intensity ratio
is constant at the 30\% level from $B$ to F160W (about the level at which
we can retrieve a value). We cannot put 
such a fine limit on the color of the components
of object 5, since it was observed on the PC chip which does not
include enough objects to register the coordinates accurately at the
sub-pixel level against the NICMOS data, which would be required to
measure color-introduced shifts given its component separation
of $0 \farcs 25$. As far as we can tell from these data on close pairs of
Lyman $\alpha$ emitters, both components have similar star-forming
properties and histories, as revealed shortward of 5000 \AA\
in their emitted frames.

\section{CONCLUSIONS}

Comparing NICMOS and IRTF observations in the emitted optical range to
WFPC2 data in the emitted ultraviolet, we have extended our probe of
the nature and evolutionary status of the 53W002 galaxy grouping
at $z=2.4$, considering both the three AGN in our field and accompanying
star-forming objects selected for narrow Lyman $\alpha$ emission.

It is useful to compare these low-luminosity objects to the Lyman break 
galaxies at $z>3$ observed in [O III] by Teplitz et al. (2000), most 
with narrowband filters in preselected $z$ ranges. 
We constructed hybrid equivalent widths for
[O III] $\lambda 5007$ -- from their spectroscopic
fluxes and imaging broad-band magnitudes -- in the
emitted frame. Their sample is more luminous in the 5000-\AA\   
continuum than ours, as might be expected from the selection methods.
We can compare this spectral region directly between the two samples, 
since it falls in the $K$ band for 
$z=3.3$ and the $H$ or NICMOS F160W band at $z=2.4$.
Lack of detected H$\beta$ coupled with the observed [O III]
strengths lead Teplitz et al. to infer [O/H] = 0.2--0.9 $\times$ 
solar for their objects, concluding that
``most LBGs appear likely to have less than solar metallicity, and yet not to 
lie in the extremely low Z regime seen in low-mass local galaxies".
We argue that the less luminous, and bluer, Lyman $\alpha$ emitters
do extend into this regime. Their 5000-\AA\   luminosities, averaged
logarithmically, are typically about three times lower than the Lyman-break 
systems, before any differential reddening correction (which would
only increase this difference). Furthermore, we see evidence for
a luminosity-metallicity connection within our sample, since the
bluest objects are also the faintest in the emitted ultraviolet,
as seen in Figure 5.
We might speculate that either the more luminous objects began star formation
earlier, or that they have deeper potential wells and therefore
can self-enrich their gas supply more efficiently than the less
luminous, presumably lower-mass objects we see at $z=2.4$.





Comparing this set of AGN and star-forming objects in a narrow
redshift range gives some hints on how AGN hosts and other
systems differ at this epoch. The AGN are interacting with
their surroundings, as shown by structures that may be reflection and 
ionization cones. These are extensive enough to suggest material ejected from
the AGN hosts, driven either by the AGN themselves or by associated
star formation. In either case, abundances in these winds,
which are strong [O III] as well as Lyman $\alpha$ sources, 
are likely to be several times higher than in the star-forming objects,
suggesting that star formation began earlier and perhaps at
a higher rate in the AGN hosts. In contrast, the abundances in the
fainter objects with active star formation are substantially
below solar, so that either they have only recently begun star
formation or have such shallow potential wells that they do not
self-enrich. Either case poses interesting questions for the
number density and fate of these objects - if they have been
forming stars for only a few times $10^7$ years, why do we see so
many of them? On the other hand, if they are losing material so rapidly, they
should fade by a large factor unless replenished by infalling gas
against the trend of mass loss driven by star formation. Multiple
episodes of infall, star formation, and outflow might be needed
if these objects are the precursors of luminous present-day galaxies.

If these systems are indeed losing their ISM by global winds,
star formation will shut down quickly and they will fade
passively. From the Bruzual \& Charlot (1993) models,
such passive evolution will amount to about 4--6 magnitudes in $V$,
depending on just how long the initial burst lasts. That would
give these objects present-epoch $V$ luminosities of order 
$2 \times 10^{37} - 5 \times 10^{38}$ erg s$^{-1}$, for
absolute magnitudes $M_V = -14$ to $-17$, respectable values
for current gas-poor dwarf galaxies. It is therefore possible
that in these systems we are observing an epoch of sweeping
driven by starbursts, as described by Wyse \& Silk (1985)
and related to the nearby systems observed by Marlowe et al. (1995).
By contrast, the AGN may have hosted multiple or very protracted bursts
of star formation, while the subgalactic objects
have undergone fewer bursts, resulting in lower O/H, and perhaps
winds which are overall less energetic (thought still sufficient to escape
their shallower potential wells).

Taken together, these results make sense in a broad hierarchical
scheme, with the AGN forming in those systems which started
from the strongest fluctuations, collapsed first, and underwent
the earliest star formation, leaving the lower-mass objects
to collapse later, perhaps undergoing their first widespread
star formation near the observed epoch at $z=2.4$. However,
the unresolved questions about the fate of the star-forming
objects still make it difficult to tie these systems together with
present-day galaxies.

\acknowledgments

This work was supported by NASA through STScI grants GO-07459.*-96A.
Luminosities of nearby star-forming systems were calculated using IUE data
retrieved from the MAST archive maintained at STScI. Ned Wright's
WWW calculator for cosmological quantities was a time-saver, particularly
in comparing data on objects at higher redshifts. We acknowledge, as well,
a useful conversation with Eric Bell in comparing experience fitting the $\beta$
slope in various ways. We thank the referee, Rodger Thompson, for comments
which led to improvement in our analysis as well as discussion.

\clearpage

\figcaption
{HST WFPC2 $I$-band (F814W) image of the 53W002 grouping, rotated
to put north along the $+y$ axis and supplemented with a ground-based
image in the area outside the WFPC2 coverage, with the fields of the
NIC3 observations marked. Numbered objects have been confirmed
to have Lyman $\alpha $ emission at $2.04 < z < 2.5$, except for
object number 1 whose line identification is tentative. Objects 6, 18, and 19 
are spectroscopically classified AGN (number 6 is the radio galaxy 53W002 
itself). Two of the candidates from Pascarelle et al. (1996, their
objects 10 and 13) within the
NICMOS fields have since been found to be stars, marked here with
surrounding squares. The field shown spans 94$\times$111 arcseconds. 
The deepest Lyman $\alpha$ search includes
the WFPC2 field, but not the area outside it.
\label{fig1}}

\figcaption
{Drizzled NIC3 images in F160W (1.6$\mu$m) for the two overlapping 
fields observed, displayed with a linear intensity mapping. Each field 
subtends $54\times 51$ arcseconds. The instrumental ($x,y$) coordinates are 
oriented close to cardinal directions, as shown
in Fig. 1. For the northern field, the instrumental $+x$-axis, which
runs vertically as displayed here, points $3.9^\circ$ west
of north, while for the southern field it points $0.6^\circ$ east
of north. The three
dithered subpointings for each field were offset 8.3 NIC3 pixels 
along the $x$-axis (approximately N-S) from
one another. High-redshift objects from Fig. 1 are marked. All three
AGN exhibit resolved structure, dominated by redshifted [O III] emission.
For objects 6 and 19, these coincide with ionization/reflection cones
seen in the UV emission lines, while for object 18 the structure is
brighter and more complex. For orientation of the two fields on the
sky, the same objects appear at the lower left corner of the northern
field and the upper right corner of the southern field.
\label{fig2}}

\figcaption
{Matching sections of the stacked IRTF images in the CVF
passband including H$\alpha$+[N II] (left) and broadband K (right).
North is at the top and east to the left, across a $43 \times 58$-arcsecond area.
The images are approximately scaled in intensity so that a pure
continuum object has the same brightness in both. The strong
emission in object 6 stands out, and is seen at a less impressive
level in 18 and 19.
\label{fig3}}

\figcaption
{Comparison of the spectral energy distributions (SEDs),
as used in fitting the slope parameter $\beta$. The three AGN,
at the top in flux, are shown as connected open circles, while
the fainter star-forming objects are shown with connected filled 
circles. Since some of these objects are at redshifts
different from the $z=2.39$ of the 53W002 grouping, the
fixed passbands introduce a visible dispersion in the effective emitted
wavelengths of their flux data. Comparison of various
photometric recipes suggests typical errors of 0.15 magnitude
for the three shorter-wavelength WFPC2 bands and 0.30 magnitude
for the two NICMOS bands, largely driven by uncertainties in
the curve of growth for slightly resolved objects; individual error
bars are not plotted to reduce clutter.
For comparison, we overlay a smoothed composite IUE/optical integrated
spectrum of the blue compact dwarf galaxy UGC 9560 as presented
by McQuade, Calzetti, \& Kinney (1995).
\label{fig4}}

\figcaption
{Relations between UV luminosity L(1200) and
spectral index $\beta$ (top) and spectroscopically measured O/H
(bottom). Filled circles show local objects, with well-known
and extreme examples named, while open circles are
star-forming objects in the 53W002 field. The vertical pairs of open circles
in the upper panel indicate values as observed and the amount of
reddening correction from the Meurer et al. (1999) prescription.
In both diagrams, the high-redshift objects extend into ranges of
parameter space including a low metallicity/high-luminosity
combination not found in the local comparison sample.
Since only limits to the metallicity are available from $\beta$,
the whole sample of high-redshift objects is represented by
an open-ended box in the abundance panel.
\label{fig5}}

\figcaption
{Slices through the NICMOS grism spectra at the expected locations of [O III]
$\lambda 5007$, showing detections in the two star-forming objects 11 and 12
at $z=2.05,2.45$. These are $0.\ \farcs 6$-width extractions, with scattered
light from nearby objects subtracted using a 25-pixel boxcar filter.
The data for object 11 have been offset vertically, with the zero-flux
line plotted.
The expected location of the stronger [O III] line $\lambda 5007$
is marked in each case.
Emission is detected from these two objects at about the $3 \sigma$ level, 
although at low levels the statistics are not quite Gaussian. The
implied line fluxes are $4-6 \times 10^{-15}$ erg cm$^{-2}$ s$^{-1}$
\AA\ $^{-1}$, for equivalent widths 100--130 \AA\  in the emitted
frame using the broadband images for continuum flux.  
\label{fig6}}

\figcaption
{NICMOS grism spectra of the three AGN in the 53W002 grouping,
covering the regions around redshifted H$\beta$+[O III]. Much of the
apparent line width in 53W002 itself and object 18 is due to
spatially resolved emission, as seen in Lyman $\alpha$ and H$\alpha$.
The spectra of objects 6 and 18 have been offset vertically by 0.002 and 0.005
counts/second respectively, with their zero-flux lines plotted.
The location of [O III] $\lambda 5007$ at the redshift of
Lyman $\alpha$ is shown by vertical lines.
The [O III] peaks, especially in object 18, are
shifted from the expected wavelength based on the redshift of
Lyman $\alpha$ (vertical lines) by spatially resolved structure.
\label{fig7}}

\figcaption
{Emission-line structure of the three AGN at $z=2.39$. Each box
spans 3 arcseconds on a side, all oriented approximately with north upward
(NICMOS images are at the instrumental orientation to avoid resampling).
Two [O III] images are shown. The ``direct" one is from the F160W 
drizzle-reconstructed images, using the F110W data and a typical scale
factor for approximate continuum subtraction. These images may have 
an improperly subtracted continuum residual at the peak. The ``spectrum" image
is from the slitless grism data, with the continuum subtracted
by interpolation in the dispersion direction (which runs vertically here). 
The Lyman $\alpha$
images are from the F410M WFPC2 observations described in P96 and P98,
and the H$\alpha$ data are from the IRTF data with continuum 
subtraction using a scaled $K$-band image. Much of the most extended
line emission from 53W002 and object 19 is of low enough surface
brightness to be most prominent in the F160W-F110W direct image.
The images of object 19 are not centered on its core to show this structure
to the south-southeast. For object 18, the total line flux is dominated
by the ``cloud" northeast of the continuum core, and the core itself
is more prominent in both [O III] renditions than in Lyman $\alpha$. 
\label{fig8}}

\figcaption
{High-redshift objects which appear as close doubles in optical
images. The $B$ images are shown both in the original (WF interleaved or
PC) $0 \farcs 045$ pixel scale and after resampling and convolution to
match the F160W PSF, and with the corresponding regions in the
drizzled F110W and F160W images with $0 \farcs 1$ pixels. Each region is
$2 \farcs 5$ on a side, aligned with the NICMOS detector coordinates
so that north is roughly to the right (as depicted in Fig. 1). 
Consistency of the image centroids indicates that the color of the
two components must in each case be similar across the 0.4-1.6$\mu$m
range.
\label{fig9}}

\clearpage

\begin{table*}
\begin{center}
\begin{tabular}{rcccl}
\tableline
\tableline
P96 ID  & $\alpha, \delta$ (J2000) & Other ID &  $z$ & Notes\\
 6   & 17 14 14.74 +50 15 28.9 & 53W002  &  2.390 & \\
18   & 17 14 11.89 +50 16 00.5 &  AGN    &  2.393 & detached companion at $z=2.406$\\
19   & 17 14 11.27 +50 16 08.7 &  AGN    &  2.397 & \\
12   & 17 14 11.66 +50 15 49.3 &         & 2.053 & double?\\
11   & 17 14 12.63 +50 15 36.8 &         & 2.451 & double \\
 5   & 17 14 13.54 +50 15 34.7 &         & 2.048 & double? \\
60   & 17 14 08.16 +50 15 52.8 &         &   2.30: & \\
61   & 17 14 08.53 +50 15 52.2 &         &  unknown & \\
 1   & 17 14 10.53 +50 15 09.8 &         &   3.1: & \\
10   & 17 14 08.17 +50 16 01.4 & star    & ... & \\
13   & 17 14 10.98 +50 15 54.3 & star    & ... & \\
\tableline
\end{tabular}
\tablenum{1}
\caption{
Candidate objects and redshifts \label{tbl1}}
\end{center}
\end{table*}

\clearpage

\begin{table*}
\begin{center}
\begin{tabular}{lccrcrc}
\tableline
\tableline
Name & $K$ & F(H$\alpha$+[N II])\tablenotemark{a} &  EW &   F([O III]) &  EW &  
F(Ly$\alpha$) \\
W02  & 18.55  &  1.7e-15   &     345 & 2.5e-15  &  701   &  2.8e-16 \\
18   & 19.39  &  6.3e-16   &     221 & 2.2e-15  &  600   &  4.7e-16 \\
19   & 18.55  &  4.0e-16   &      81 & 8.9e-16  &   89   &  8.0e-16 \\
     &        &            &         &          &        &          \\
Cyg A &       &   1.7e-13  &     693 & 1.6e-13  &  678   &          \\     
\tableline
\end{tabular}
\tablenotetext{a}{Line fluxes are in erg cm$^{-2}$ s$^{-1}$ \AA\ $^{-1}$,
and equivalent widths in \AA\  in the emitted frame.}
\tablenum{2}
\caption{
Infrared emission lines in AGN \label{tbl2}}
\end{center}
\end{table*}

\clearpage

\begin{table*}
\begin{center}
\begin{tabular}{lcccccccc}
\tableline
\tableline
Object & $z$ & F$_{\rm B}$\tablenotemark{a} &   F$_{\rm V}$     &  
F$_{\rm I}$   &     F$_{110}$ & F$_{160}$ &  F$_{175}$ & F$_{\rm K}$  \\
01  & 3.1:\tablenotemark{b}   & 0.74 & 0.62 &  0.39 & -0.05  & -0.49  & ...   & 0.49 \\ 
05  & 2.048  & 0.90 & 0.75 &  0.48 &  0.29  &  0.14  & ...   & ...  \\
11  & 2.451  & 0.78 & 0.62 &  0.40 &  0.15  &  0.21  & ...   & ...  \\
12  & 2.053  & 0.98 & 0.75 &  0.61 &  0.37  &  0.05  & 0.03  & ...  \\
60  & 2.30:  & 0.64 & 0.38 &  0.11 &  0.09  & -0.42  & 0.18  & 0.50 \\
61  &  ...   & 0.28 & 0.08 &  0.11 & -0.45  & -0.36  & ...   & ...  \\
AGN:&        &      &      &       &        &        &       &      \\
06  & 2.390  & 1.22 & 1.25 &  1.19 &  1.10  &  1.23  & ...   & 1.16 \\
18  & 2.393  & 1.29 & 1.10 &  0.93 &  0.85  &  1.12  & 0.94  & 0.83 \\
19  & 2.397  & 1.31 & 1.13 &  1.00 &  1.03  &  1.10  & 1.02  & 1.06 \\
\tableline
\end{tabular}
\tablenotetext{a}{Fluxes are in log (F$_\lambda /10^{-19}$ erg cm$^{-2}$
s$^{-1}$ \AA\ $^{-1}$).}
\tablenotetext{b}{Redshift uncertain from one possible feature, assumed 
Lyman $\alpha$}
\tablenum{3}
\caption{
Broad-band photometry of star-forming objects \label{tbl3}}
\end{center}
\end{table*}

\clearpage

\begin{table*}
\begin{center}
\begin{tabular}{lc}
\tableline
\tableline
Object & $\beta$\cr
01 &  $-2.26 \pm 0.14$\cr
05 &  $-1.41 \pm 0.09$\cr
11 &  $-1.15 \pm 0.11$\cr
12 &  $-1.62 \pm 0.11$\cr
60 &  $-1.77 \pm 0.17$\cr
61 &  $-1.27 \pm 0.28$\cr
AGN: & \omit \cr
06 &  $-0.08 \pm 0.14$\cr
18 &  $-0.44 \pm 0.22$\cr
19 &  $-0.44 \pm 0.23$\cr
\tableline
\end{tabular}
\tablenum{4}
\caption{
Fitted continuum slopes $\beta$ \label{tbl4}}
\end{center}
\end{table*}

\clearpage
\begin{table*}
\begin{center}
\begin{tabular}{lcc}
\tableline

Object & L(1200)\tablenotemark{a} &  L(5000)\\
1\tablenotemark{b} & 3.1e40 & 3.5e39 \\ 
5  & 8.9e40  & 9.7e39 \\
11 & 4.7e40  &  1.8e40 \\
12 & 4.3e40  &  8.0e39 \\
60 & 3.7e40  &  4.1e39 \\
\tableline
\end{tabular}
\tablenotetext{a}{Luminosities are in erg s$^{-1}$ \AA\ $^{-1}$.}
\tablenotetext{b}{Assumed at $z \approx 2.4$}
\tablenum{5}
\caption{
Derived UV and optical continuum luminosities \label{tbl5}}
\end{center}
\end{table*}
\clearpage

\begin{table*}
\begin{center}
\begin{tabular}{lcccrr}
\tableline
\tableline
Object & Ly$\alpha$\tablenotemark{a} & [O III] $\lambda 5007$ &  H$\beta$
&  EW(5007)\tablenotemark{b} & EW(H$\beta$) \\
1  &  2.6e-17 & $<4$e-17 & $<4$e-17 & $<360$ & $<360$ \\
5  &  1.7e-16 & $<4$e-17 & $<4$e-17  & $<95$ & $<95$ \\
11\tablenotemark{c} &  3.4e-17 & 1.9e-16 & $<8$e-17 & 340 & $<140$\\ 
12  & 3.4e-16 & 1.3e-16 & $<4$e-17 &  380 & $<120$ \\
60  & 3.1e-17 & $<4$e-17 & $<4$e-17  & $<320$ & $<320$ \\
\tableline
\end{tabular}
\tablenotetext{a}{Line Fluxes are in erg cm$^{-2}$
s$^{-1}$ \AA\ $^{-1}$.}
\tablenotetext{b}{Equivalent widths are in the emitted frame. The continuum
was derived from F160W images, since it is too weak to be measured in the
NICMOS spectra.}
\tablenotetext{c}{Superimposed on the spectrum of a brighter object.}
\tablenum{6}
\caption{
Emission-line properties for star-forming objects \label{tbl6}}
\end{center}
\end{table*}

\clearpage

\end{document}